\documentclass[letterpaper,twoside,twocolumn,english,aps]{revtex4}

\usepackage[T1]{fontenc}
\usepackage[latin9]{inputenc}
\usepackage{amsmath}
\usepackage{amssymb}
\usepackage{graphicx}
\usepackage{esint}

\makeatletter

\pdfpageheight\paperheight
\pdfpagewidth\paperwidth

\providecommand{\tabularnewline}{\\}

\@ifundefined{textcolor}{}
{%
 \definecolor{BLACK}{gray}{0}
 \definecolor{WHITE}{gray}{1}
 \definecolor{RED}{rgb}{1,0,0}
 \definecolor{GREEN}{rgb}{0,1,0}
 \definecolor{BLUE}{rgb}{0,0,1}
 \definecolor{CYAN}{cmyk}{1,0,0,0}
 \definecolor{MAGENTA}{cmyk}{0,1,0,0}
 \definecolor{YELLOW}{cmyk}{0,0,1,0}
}

\makeatother

\usepackage{babel}
\begin{document}

\title{Microscopic Origin and Universality Classes of the Efimov Three-Body
Parameter}

\author{Pascal Naidon$^{1}$, Shimpei Endo$^{2}$, and Masahito Ueda$^{2}$}

\affiliation{$\,^{1}$RIKEN Nishina Centre, RIKEN, Wak\={o} 351-0198, Japan, }

\affiliation{$\,^{2}$Department of Physics, University of Tokyo, 7-3-1 Hong\={o},
Bunky\={o}-ku, T\={o}ky\={o} 113-0033, Japan}

\date{\today}
\begin{abstract}
The low-energy spectrum of three particles interacting via nearly
resonant two-body interactions in the Efimov regime is set by the
so-called three-body parameter. We show that the three-body parameter
is essentially determined by the zero-energy two-body correlation.
As a result, we identify two classes of two-body interactions for
which the three-body parameter has a universal value in units of their
effective range. One class involves the universality of the three-body
parameter recently found in ultracold atom systems. The other is relevant
to short-range interactions that can be found in nuclear physics and
solid-state physics.
\end{abstract}
\maketitle
\begin{figure}
\includegraphics[bb=0bp 160bp 960bp 540bp,clip,scale=0.22]{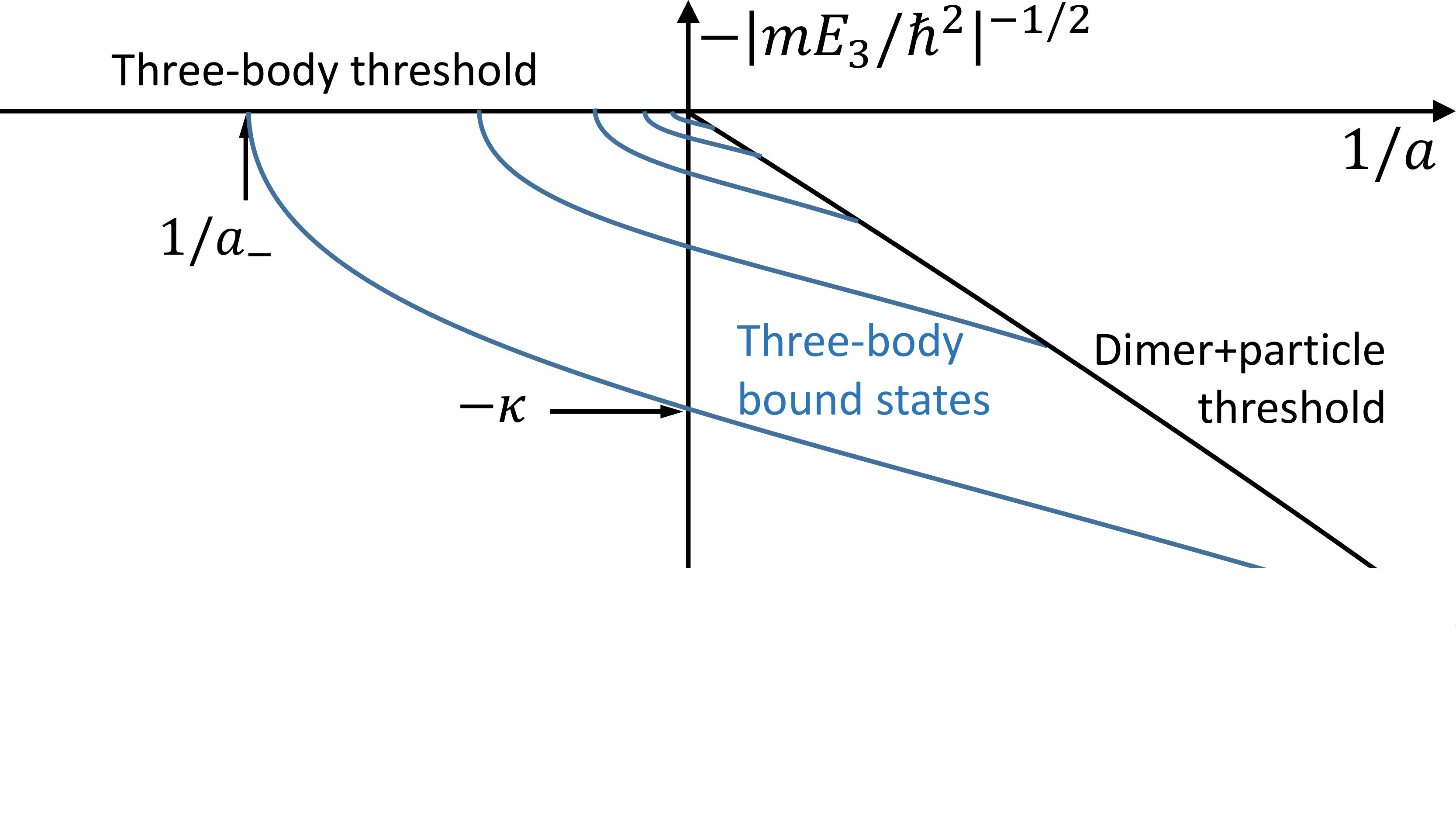}

\caption{\label{fig:Efimov-plot}Schematic Efimov plot: three-body energy $E_{3}$
scaled as an inverse length as a function of the inverse scattering
length $1/a$. The arrows indicate the scattering length $a_{-}$
at which an Efimov trimer state appears, and its binding wave number
$\kappa$ at unitarity ($a\to\infty)$, either of which serves as
a measure of the three-body parameter. }
\end{figure}

The Efimov effect is a universal low-energy quantum phenomenon, which
was originally predicted in nuclear physics~\citep{Efimov1970} and
has rekindled considerable interest since its experimental confirmation
with ultracold atoms \citep{Kraemer2006,Ottenstein2010,Huckans2009,Williams2009,Zaccanti2009,Barontini2009,Wenz2009,Pollack2009,Gross2009,Knoop2009,Lompe2010PRL,Lompe2010Science,Nakajima2010,Gross2010,Nakajima2011,Berninger2011,Ferlaino2009,Ferlaino2010,Machtey2012,Zenesini2012,Knoop2012}.
It is also expected to occur in solid-state physics~\citep{Nishida2013,Omachi2013}.
This universality stems from the effective three-body attraction that
occurs between particles interacting with nearly resonant short-range
interactions. As a result of this attraction, three particles may
bind even when the interaction is not strong enough to bind two particles.
Furthermore, an infinite series of such three-body bound states exists
near the unitary point where the interaction is resonant, i.e. where
a two-body bound state appears and the $s$-wave scattering $a$ length
diverges. The typical three-body energy spectrum for such systems
is represented in Fig.~\ref{fig:Efimov-plot} in units of inverse
length. Near zero energy and large scattering lengths, the three-body
spectrum is invariant under a discrete scaling transformation by a
universal factor $e^{\pi/s_{0}}\approx22.7$ for identical bosons,
where $s_{0}\approx1.00624$ characterises the strength of the three-body
attraction.

A notable consequence of the Efimov effect is the existence of another
physical scale beyond the two-body scattering length to fix the low-energy
properties of the system. This scale is known as the three-body parameter.
In zero-range models, it manifests itself as the necessity to introduce
a momentum cutoff or a three-body boundary condition. It can be characterised,
for instance, by the scattering length $a_{-}$ at which a trimer
appears or by its binding wave number $\kappa$ at unitarity, as indicated
in Fig.~\ref{fig:Efimov-plot}. Because of the discrete scaling invariance,
it is defined up to a power of $e^{\pi/s_{0}}$. In this Letter,
we will focus on the ground Efimov state, which slightly deviates
from the discrete-scaling-invariant structure, but is more easily
observed and computed, and still reveals the essence of the physics
behind the three-body parameter.

Three important questions can be raised concerning the three-body
parameter. Is there a simple mechanism that determines the three-body
parameter from the microscopic interactions? What is the microscopic
length scale which determines the three-body parameter? Finally, if
there is such a length scale, what are the conditions for the three-body
parameter to be related to that length scale through a universal dimensionless
constant, as suggested by experimental observations~\citet{Gross2010,Berninger2011}
and recent calculations~\citet{JiaWang2012}? This Letter answers
these three questions for systems of identical bosons (or three distinguishable
fermions with equal mass) where the resonant interaction can be described
by a single scattering channel. In ultracold atoms experiments, the
interaction is made resonant by using magnetic Feshbach resonances~\citep{Chin2010}.
The present results are thus applicable to the case of broad Feshbach
resonances, which are dominated by their open channel, but not to
narrow Feshbach resonances, which are strongly affected by their closed
channel~\citep{Petrov2004}.

The question of the physical mechanism setting the three-body parameter
was addressed in Refs.~\citep{JiaWang2012,Naidon2012b} for van der
Waals interactions, which decay as $1/r^{6}$ at large interparticle
distance $r$. The numerical investigation of Ref.~\citep{JiaWang2012}
found that in the hyperspherical formalism, in addition to the three-body
Efimov attraction at large distances, a three-body repulsion appears
at short distances. The distance at which this repulsion appears is
comparable to the size of the van der Waals tail of the potential,
thus preventing the system from probing the details of the interaction
at shorter distances. Therefore, the value of the three-body parameter
is set by the van der Waals length associated with that tail. The
authors of Ref.~\citep{JiaWang2012} remarked that this three-body
repulsion is not explained by quantum reflection, as originally suggested
in Ref.~\citep{Chin2011}, but attributed it to an increase in kinetic
energy due to the squeezing of the hyperangular wave function into
a smaller volume caused by the suppression of two-body probability
inside the well or the repulsive core of the two-body potential. This
point was confirmed and clarified in Ref.~\citep{Naidon2012b} where
the kinetic energy was shown to originate from an abrupt change of
the geometry of the three-particle system caused by the two-particle
exclusion in the van der Waals region. At large separation, the system
has indeed an elongated geometry due to its Efimov nature, but it
must deform to an equilateral configuration to accommodate for the
mutual exclusion between the particles. Reference~\citep{Naidon2012b}
showed that this deformation causes a nonadiabatic increase in kinetic
energy that manifests itself as a three-body repulsive barrier. This
phenomenon could be reproduced by simple models involving only the
knowledge of the pair correlation causing the mutual exclusion between
two particles at short separations.

One may wonder whether these findings extend to other physical systems.
Indeed, the same deformation mechanism is expected to occur in systems
for which the two-body interactions tend to suppress the two-body
probability at short distance. Thus, pair correlation should provide
the essential information that determines the three-body parameter
and energy of the three-body system.

To investigate the role of pair correlation, we use a simple model
that reproduces the pair correlation and can be solved exactly for
three particles, and then compare it with full exact calculations.
We cannot use a zero-range model  because such a model would reproduce
the asymptotically free part of the two-body wave function (i.e. the
on-shell $T$-matrix elements), but not its short-range correlation
(i.e. the off-shell $T$-matrix elements). We thus follow the approach
introduced in Ref.~\citep{Naidon2012b}, where the interaction is
modelled by a separable potential \citep{Yamaguchi1954}, $\hat{V}=\xi\vert\chi\rangle\langle\chi\vert$,
which retains much of the mathematical simplicity of a contact potential,
while enabling us to reproduce any pair correlation at zero energy.
Indeed, for a given zero-energy two-body $s$-wave radial wavefunction
$u_{0}(r)$ with the asymptotic limit $1-\frac{r}{a}$, where $a$
is the scattering length, one can construct a separable potential
reproducing this wave function exactly by choosing the following form
(in momentum representation):
\begin{equation}
\chi(q)=1-q\int_{0}^{\infty}dr\left(1-\frac{r}{a}-u_{0}(r)\right)\sin(qr),\label{eq:Chi}
\end{equation}
\begin{equation}
\xi=4\pi\left(\frac{1}{a}-\frac{2}{\pi}\int_{0}^{\infty}dq\vert\chi(q)\vert^{2}\right)^{-1}.\label{eq:Xi}
\end{equation}
\begin{table}
\begin{tabular}{lllllll}
\hline 
Potential & $g_{3}$ & $g_{3}^{\prime}$ & $a_{-}$ & $a_{\ensuremath{-}}^{\prime}$ & $E_{3}$ & $E_{3}^{\prime}$\tabularnewline
\hline 
\hline 
{\footnotesize{Yukawa}} & 1.35 & 1.38 & $-5.73$ & $-6.55$ & $-0.172$ & $-0.134$\tabularnewline
{\footnotesize{Exponential}} & 1.17 & 1.16 & $-10.7$ & $-11.0$ & $-0.047$ & $-0.042$\tabularnewline
{\footnotesize{Gaussian}} & 2.12 & 2.14 & $-4.27$ & $-4.47$ & $-0.236$ & $-0.223$\tabularnewline
{\footnotesize{Morse ($r_{0}\!=\!1$)}} & 0.294 & 0.295 & $-12.3$ & $-12.6$ & $-0.0325$ & $-0.0299$\tabularnewline
{\footnotesize{Morse $(r_{0}\!=\!2)$}} & 0.205 & 0.205 & $-16.4$ & $-16.3$ & $-0.0174$ & $-0.0166$\tabularnewline
{\footnotesize{Pöschl-Teller $(\alpha\!\!=\!\!1)$}} & 0.797 & 0.802 & $-6.02$ & $-6.23$ & $-0.135$ & $-0.123$\tabularnewline
\hline 
\end{tabular}

\caption{\label{tab:Potentials}Three-body properties obtained for various
potentials considered in Ref.~\citep{Moszkowski2000}, where $g_{3}$
denotes the factor required to multiply the potential so that the
ground-state Efimov trimer appears at the three-body threshold, $a_{-}$
denotes the scattering length for that factor, and $E_{3}$ denotes
the energy of the ground-state Efimov trimer at unitarity ($a\to\infty$).
The symbols without a prime indicate that the values are taken from
Ref.~\citep{Moszkowski2000}, and those with a prime show our results
based on the pair correlation using the separable model given by Eqs.~(\ref{eq:Chi})
and (\ref{eq:Xi}). The same units as in Ref.~\citep{Moszkowski2000}
are used.}
\end{table}

\begin{figure}
\includegraphics[scale=0.7]{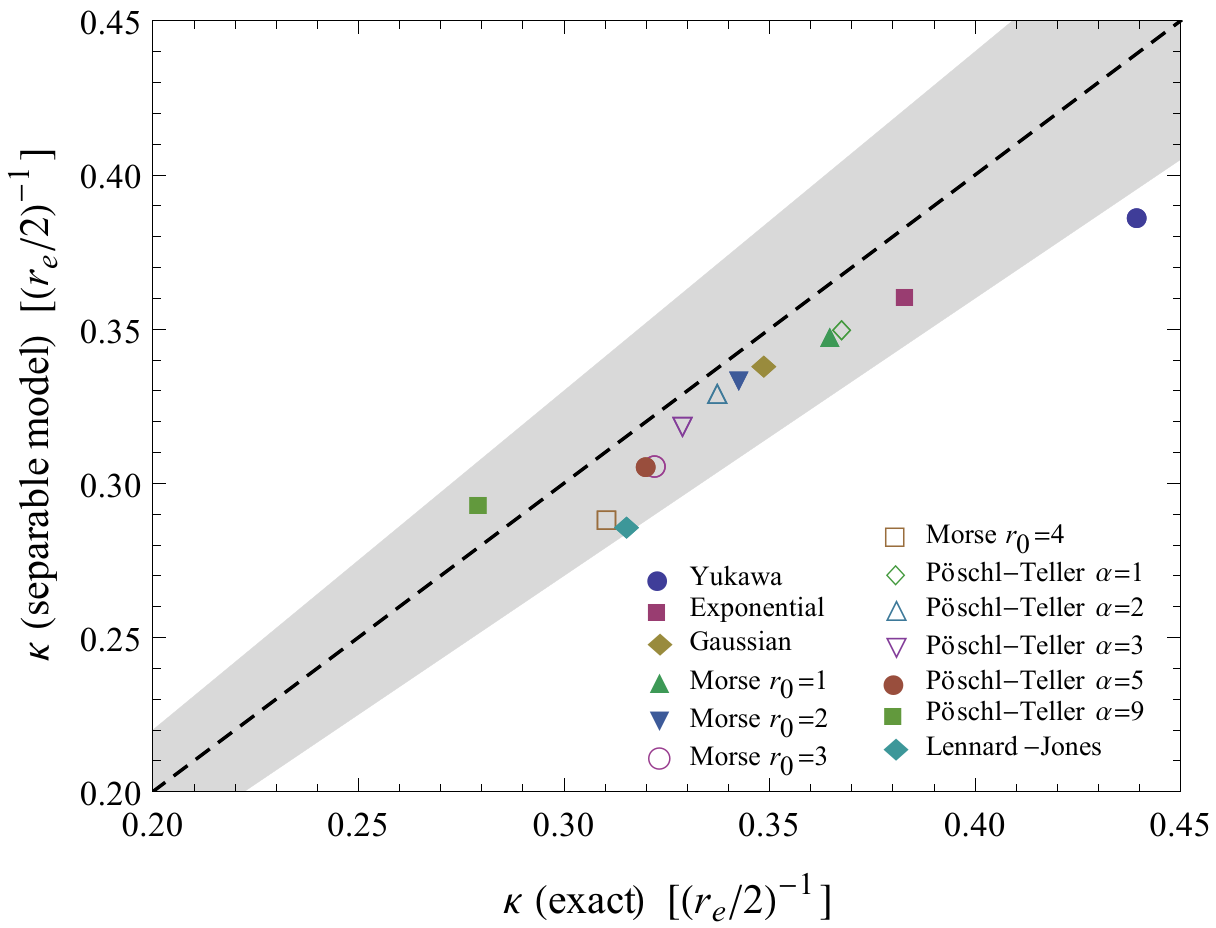}

\caption{\label{fig:Model}Binding wave number $\kappa$ of the ground-state
trimer at unitarity (see Fig.~\ref{fig:Efimov-plot}) calculated
from the separable model in Eqs.~(\ref{eq:Chi}) and (\ref{eq:Xi})
using the zero-energy pair wave function $\psi_{0}$, versus its exact
value, for various two-body potentials. The exact values are taken
from Ref.~\citep{JiaWang2012} for the Lennard-Jones potential (with
only one bound state) and from Ref.~\citep{Moszkowski2000} for all
the other potentials. The binding wave number is expressed in units
of the effective range $r_{e}$ of each potential, which is calculated
exactly from Eq.~(\ref{eq:EffectiveRange}). The shaded area represents
the region of 10\% or less deviation from the exact results.}
\end{figure}

This simple prescription reproduces the low-energy two-body physics,
in particular the two-body bound state around the unitary limit $a\to\infty$.
We construct separable potentials that reproduce the pair correlation
of the various two-body potentials considered in Refs.~\citep{JiaWang2012}
and \citep{Moszkowski2000}. The three-body problem for a separable
interaction can be cast in the form of an integral equation in momentum
space that can easily be solved numerically~\citep{Lee2007,Naidon2011,Naidon2012b}.
The results are shown in Table~\ref{tab:Potentials} where we indicate
the values $a_{-}$ and $\kappa$ for the ground-state trimer. They
agree with the exact calculations of Refs.~\citep{JiaWang2012} and
\citep{Moszkowski2000} to within a few percents for each of these
potentials. This can be checked in Fig.~\ref{fig:Model} where the
value of $\kappa$ in our model is plotted against its exact value.
The method presented here therefore appears as a simple and efficient
way to estimate the three-body parameter, and more generally low-energy
properties for various kinds of interaction potentials.

Now that we have established the connection between the pair correlation
and the three-body parameter, we are in a position to ask which length
scale in the pair correlation determines the three-body parameter.
For most physical interactions, the major effect of pair correlations
is to suppress probability at short distance with respect to the free
wave. As discussed previously, this creates a three-body repulsion
through the nonadiabatic deformation effect. Although the precise
shape of the repulsive barrier depends on the particular two-body
potential, it should be the length scale associated with the two-body
suppression that sets the location of the three-body repulsion, and
therefore the three-body parameter. This length scale is given by
half the effective range $\frac{1}{2}r_{e}$~%
\footnote{The presence of the the factor $\frac{1}{2}$ is due to the conventional
definition of the effective range.%
}, which is the average size of the deviation between the asymptotic
and fully correlated probability densities~\citep{Smorodinsky1947,Bethe1949}:

\begin{equation}
\frac{1}{2}r_{e}=\int_{0}^{\infty}dr\left(\left(1-\frac{r}{a}\right)^{2}-u_{0}(r)^{2}\right)\label{eq:EffectiveRange}
\end{equation}

Thus for common interactions which tend to suppress the two-body probability
within their range, $r_{e}$ is positive and the three-body parameter,
expressed in the dimension of length, is on the order of $\frac{1}{2}r_{e}$.
Note that the effective range is commonly featured as a term in the
low-energy expansion of the scattering phase shift (i.e. the on-shell
$T$-matrix elements). However, we expect that it is not possible
to find a connection between the three-body parameter and the effective
range from a method which introduces the effective range in this manner,
since this expansion concerns only on-shell scattering and does not
describe the short-range correlation explicitly~%
\footnote{Such an approach can be applied, however, to the case of narrow Feshbach
resonances (see Ref. {[}26{]}) for which the mechanism setting the
three-body parameter is different than that of single-channel resonances
considered here. %
}.

\begin{figure}
\includegraphics[scale=0.8]{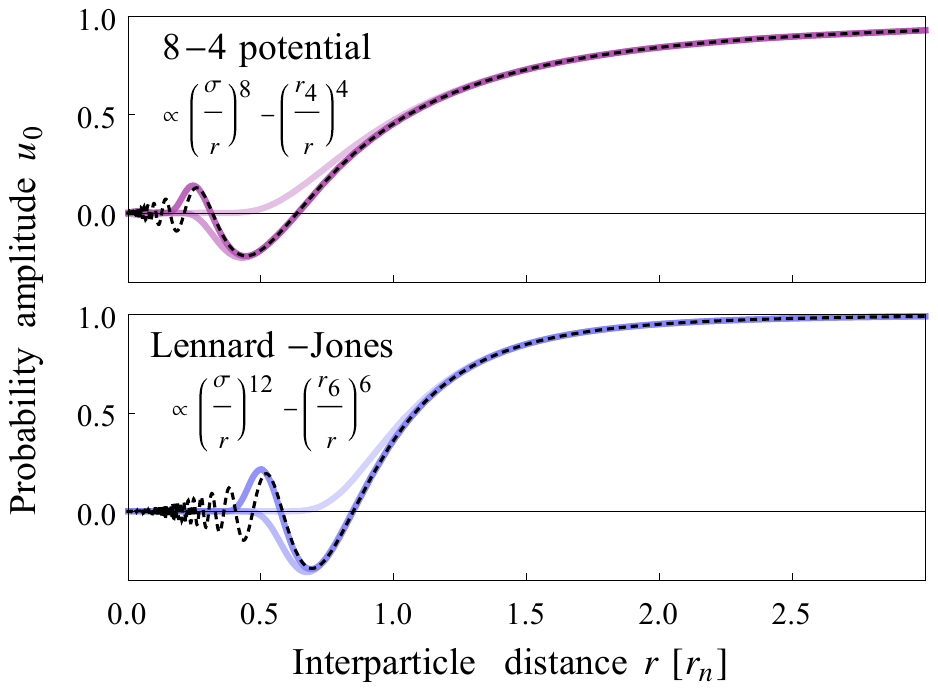}\caption{\label{fig:TwoBodyPowerLaw}Pair correlation at unitarity for potentials
decaying as power laws $-1/r^{n}$. Top: the 8-4 potential. Bottom:
the Lennard-Jones potential. In each graph, the solid curves correspond
in order of opacity to potential depths supporting respectively 1,
2, and 3 $s$-wave bound states, which are obtained by adjusting the
value of $\sigma$. The dashed curve represents the universal pair
correlation in Eq.~(\ref{eq:UniversalPairCorrelation}).}
\end{figure}

\begin{figure}
\includegraphics[bb=0bp 0bp 280bp 228bp,clip,scale=0.83]{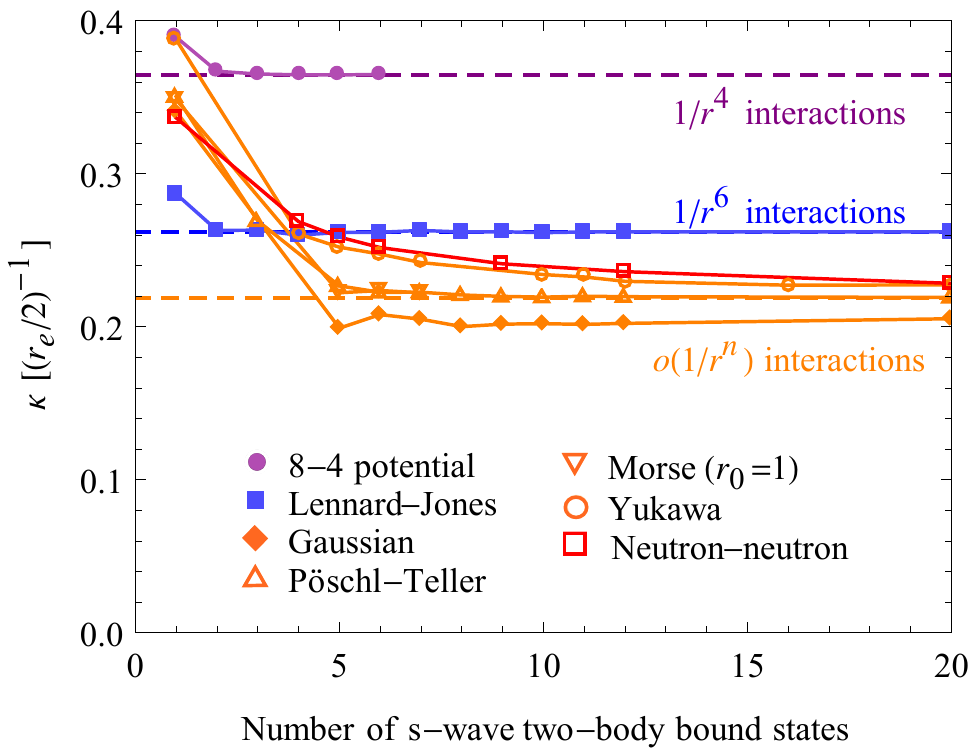}\caption{\label{fig:BindingWaveNumber}Binding wave number $\kappa$ of the
ground-state Efimov trimer calculated from the separable model in
Eqs.~(\ref{eq:Chi}) and (\ref{eq:Xi}) for pair correlations corresponding
to different two-body interactions, as a function of the depth of
these potentials as measured by the number of $s$-wave two-body bound
states. The dashed lines indicate from top to bottom the values obtained
for the universal pair correlation in Eq.~(\ref{eq:UniversalPairCorrelation})
with $n=4$ and $n=6$, and the universal pair correlation in Eq.~(\ref{eq:UniversalPairCorrelation2}),
respectively. This figure shows how the three-body parameter converges
differently and to different limits depending on the class of two-body
interaction.}
\end{figure}

We can now address the final question of whether there are classes
of interactions for which the low-energy three-body physics is universally
determined. It is clear that if the pair correlation is the same for
a certain class of potentials, they must lead to the same three-body
parameter. This is indeed the case for potentials with a power-law
decaying tail $-C_{n}r^{-n}$, such as the van der Waals tail $-C_{6}r^{-6}$
relevant to the interaction between ground-state atoms. It is well
known that the two-body wave functions in the tail of these potentials
are universally described in terms of the length scale $r_{n}=(\frac{1}{n-2}\sqrt{mC_{n}}/\hbar)^{2/(n-2)}$
. If most of the probability amplitude is located in the tail region,
which is the case if the short-range region is strongly repulsive
or attractive, all these potentials lead to a similar zero-energy
pair correlation that is known analytically:
\begin{multline}
u_{0}(r)=\Gamma(\frac{n-1}{n-2})\sqrt{x}J_{\frac{1}{n-2}}\!\!\left(2x^{-\frac{n-2}{2}}\right)\\
-\frac{r_{n}}{a}\Gamma(\frac{n-3}{n-2})\sqrt{x}J_{-\frac{1}{n-2}}\!\!\left(2x^{-\frac{n-2}{2}}\right),\label{eq:UniversalPairCorrelation}
\end{multline}
where $\Gamma$ and $J_{\alpha}$ denote the gamma and Bessel functions,
and $x=r/r_{n}$. The universality of the pair correlation is illustrated
in Fig.~\ref{fig:TwoBodyPowerLaw} for the $8-4$ and Lennard-Jones
$(12-6)$ potentials of various depths. This gives a simple explanation
of the observed universality of the three-body parameter in atomic
systems ranging from light helium \citep{Naidon2012a,Knoop2012} to
heavy atoms under broad magnetic Feshbach resonances~\citep{Gross2010,Berninger2011}.
Figure~\ref{fig:BindingWaveNumber} shows the binding wave number
$\kappa$ of the ground-state trimer for these power-law decaying
potentials, evaluated using our separable potential method. For potential
depths supporting more than one two-body bound state, the ground-state
trimer is in fact a resonance in the particle-dimer continuum, but
it manifests itself simply as a bound state in our model~%
\footnote{A separable potential has only one two-body bound state. In our model,
this state describes the two-body bound state of the original potential
which is close to the two-body scattering threshold, while the other
bound states of the potential are not included in our model. As a
result, the trimer has no lower-lying state to dissociate to, and
it appears as a true bound state, rather than a resonance. %
}. One can see that the value of $\kappa$ remains close to the one
obtained from the universal pair correlation. In the particular case
of a van der Waals tail, we obtain $a_{-}=-10.86(1)\, r_{6}$ and
$\kappa=0.187(1)\, r_{6}^{-1}$ in good agreement with Ref.~\citep{JiaWang2012}
and experimental observations \citep{Berninger2011,Ferlaino2011}.
Since the effective range is related to the van der Waals length $r_{6}$
through $r_{e}=\frac{4\pi}{3\Gamma(3/4)^{2}}r_{6}\approx2.78947r_{6}$,
these results correspond to $a_{-}=-7.78(1)\times(\frac{1}{2}r_{e})$
and $\kappa=0.261(1)\times(\frac{1}{2}r_{e})^{-1}$. 

\begin{figure}
\includegraphics[scale=0.85]{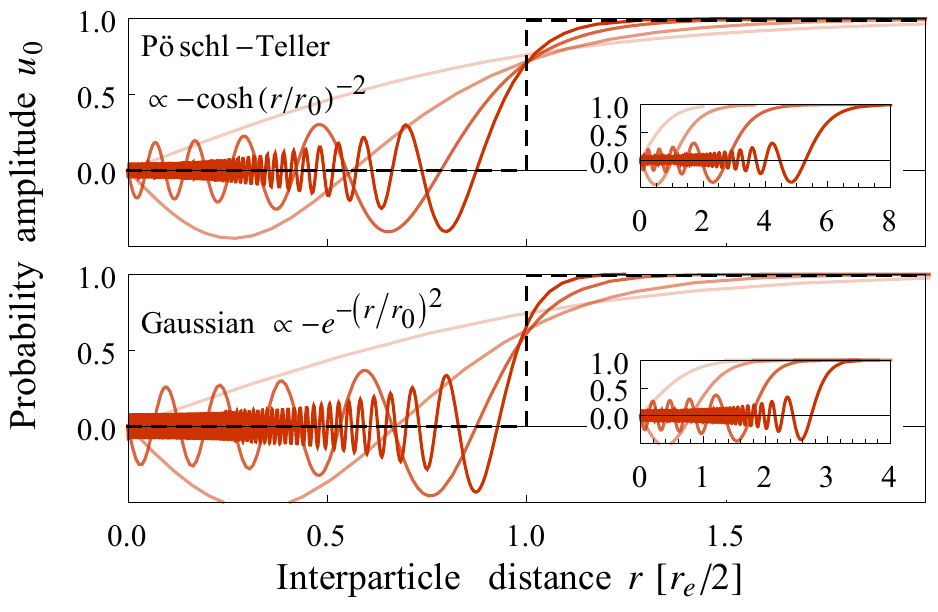}

\caption{\label{fig:TwoBodyShortRange}Pair correlation at unitarity for potentials
decaying faster than any power law. Top: the Pöschl-Teller potential;
bottom: the Gaussian potential. In each graph, the solid curves correspond
in order of opacity to potential depths supporting respectively 1,
2, 10, and 120 $s$-wave bound states. The dashed lines show the universal
pair correlation limit in Eq.~(\ref{eq:UniversalPairCorrelation2}).
The distance is scaled in units of $\frac{1}{2}r_{e}$ in the main
graphs, while it is shown in unscaled units of $r_{0}$ in the insets.}
 
\end{figure}

There is a second class of potentials, which decay faster than any
power law, such as the Yukawa potential and other typical nuclear
potentials, as well as screened potentials found in solid-state physics.
At first glance, the two-body wave functions for these potentials
do not seem to exhibit any particular universality. However, if the
potential features a deep attraction supporting many bound states,
the effective range near unitarity is large. This means that when
distances are expressed in units of $\frac{1}{2}r_{e}$, there is
a sharp drop of probability in the two-body wave function near $r=1$,
as represented in Fig.~\ref{fig:TwoBodyShortRange}. It can be shown
that this rescaled two-body wave function converges to a step function
in the limit of strongly attractive potentials~%
\footnote{The same happens for potentials with finite support, although the
effective range does not increase with the number of bound states
in this case.%
}. In this sense, the three-body parameter is universally determined
by the effective range of these potentials, and stems from the universal
pair correlation limit:
\begin{equation}
u_{0}(r)=\begin{cases}
0 & \mbox{for }r<\frac{1}{2}r_{e}\\
1-\frac{r}{a} & \mbox{for }r\ge\frac{1}{2}r_{e}
\end{cases}.\label{eq:UniversalPairCorrelation2}
\end{equation}

Figure~\ref{fig:BindingWaveNumber} shows the trimer binding wave
number $\kappa$ for some of these potentials, namely, the Gaussian
potential, the Pöschl-Teller potential with $\alpha=1$, the Yukawa
potential, the Morse potential with $r_{0}=1$~\citep{Moszkowski2000},
as well as the neutron-neutron interaction potential in the $\,^{1}S_{0}$
channel \citep{Wiringa1995}. While none of these calculations correspond
to a particular physical system, they capture the essence of the Efimov
physics occurring in the symmetric channel of nuclear systems, such
as the tritium nucleus. Each potential was scaled to reach unitarity,
corresponding to different possible depths of the potential. One can
see in Fig.~\ref{fig:BindingWaveNumber} that as the depth of the
potentials is increased, $\kappa$ converges to the value $\kappa=0.2190(1)\times(\frac{1}{2}r_{e})^{-1}$
obtained for the two-body correlation in Eq.~(\ref{eq:UniversalPairCorrelation2}).
The convergence is, however, very slow, because very deep potentials
(supporting hundreds of bound states) are required for the pair correlation
to approach Eq.~(\ref{eq:UniversalPairCorrelation2}).

Finally, one should note that there is a notable exception to these
considerations. One might think that the square-well potential, which
often lends itself to simple analytical treatments~\citep{Jensen1997},
is a useful model potential to investigate the physics of the three-body
parameter. However, it turns out to be a special case which does
not belong to the two classes discussed above. Even though it decays
faster than any power law, it does not belong to the second class
because of its absence of tail. In particular, the two-body wave function
near unitarity shows no progressive drop of probability in the well,
only steady oscillations which get faster as the depth of the well
increases, and therefore does not converge to the function in Eq.~(\ref{eq:UniversalPairCorrelation2}).
From this we conclude that this potential is not expected to reveal
any universality of the three-body parameter. 

To summarise, we have pointed out how the Efimov three-body parameter
is deeply connected to the zero-energy two-body correlation. This
allows us to identify the two-body effective range as the relevant
length scale setting the three-body parameter for the class of physical
interactions which suppress two-body probability at short distance.
However, it also shows that, unlike what was suggested in Ref.~\citep{JiaWang2012},
this suppression of two-body probability does not lead to a single
universal value of the three-body parameter in units of the effective
range. Indeed we find two qualitatively distinct subclasses of interactions
for which the value of the three-body parameter is universally determined.
One corresponds to short-range two-body potentials decaying as a power
law, relevant to atomic interactions, for which the three-body universality
stems from the two-body universality. The other corresponds to very
deep two-body potentials decaying faster than any power law, which
lead to an abrupt two-body suppression. Typical interactions in nuclear
physics decay faster than any power law but support only a few bound
states, so that their three-body parameter does not reach this universal
limit. In practice, however, one can expect the binding wave number
$\kappa$ to be in the range $0.2\sim0.4\times(\frac{1}{2}r_{e})^{-1}$
for most physical interactions, and in particular close to $0.35\times(\frac{1}{2}r_{e})^{-1}$
for nuclear interactions supporting at most one bound state, as can
be seen in Figs.~\ref{fig:Model} and \ref{fig:BindingWaveNumber}.
These conclusions are obtained for particles interacting through single-channel
two-body interactions, and would not apply in the presence of significant
three-body forces, or strongly energy-dependent resonant interactions
such as narrow Feshbach resonances~\citep{Petrov2004,Schmidt2012,Sorensen2012}. 

P. N. acknowledges support from RIKEN through the Incentive Research
Project funding. S. E. acknowledges support from JSPS (Grant No. 237049).
M. U. acknowledges support by Grants-in-Aid for Scientific Research
(Kakenhi No. 22340114 and No. 22103005) and the Photon Frontier Network
Program of MEXT of Japan.

\bibliographystyle{apsrev}

\end{document}